\documentclass[aip,rsi,twocolumn,reprint,groupedaddress,amsmath,amssymb,a4paper]{revtex4-1}

\usepackage{graphicx}
\usepackage{dcolumn}
\usepackage{bm}

\begin{document}

\title{High resolution magnetostriction measurements in pulsed
  magnetic fields using Fibre Bragg Gratings}

\author{Ramzy Daou}
\altaffiliation{based at: Hochfeld-Magnetlabor Dresden,
   Forschungszentrum Dresden-Rossendorf, 01314 Dresden, Germany.}

\author{Franziska Weickert}
\altaffiliation{based at: Hochfeld-Magnetlabor Dresden,
   Forschungszentrum Dresden-Rossendorf, 01314 Dresden, Germany.}

\author{Michael Nicklas}

\author{Frank Steglich}
 \affiliation{Max Planck Institute for the Chemical Physics of Solids, 01187 Dresden, Germany.}

\author{Ariane Haase}
\altaffiliation{based at: Hochfeld-Magnetlabor Dresden,
   Forschungszentrum Dresden-Rossendorf, 01314 Dresden, Germany.}

\author{Mathias Doerr}
 \affiliation{Institut f\"ur Festk\"orperphysik, TU Dresden, 01062 Dresden, Germany.}

\date{\today}

\begin{abstract}
We report on a new high resolution apparatus for measuring
magnetostriction suitable for use at cryogenic temperatures in pulsed
high magnetic fields which we have developed at the Hochfeld-Magnetlabor
Dresden. Optical fibre strain gauges based on Fibre Bragg Gratings are
used to measure the strain in small ($\sim 1$\,mm) samples. We describe the
implementation of a fast measurement system capable of resolving
strains in the order of $10^{-7}$ with a full bandwidth of 47\,kHz,
and demonstrate its use on single crystal samples of GdSb and GdSi.
\end{abstract}

\pacs{75.80.+q, 42.81.Pa}

\maketitle

\section*{Introduction}

Magnetostriction, the strain response of a material to an
external or internal magnetic field, is a thermodynamic tensor
quantity that is closely related to the magnetisation.\cite{doerr08}

Pulsed field facilities provide the highest available magnetic fields
for research purposes. Electrical and mechanical noise are
significantly higher than in steady fields, however, and the short
timescales mean that this noise cannot be effectively averaged. For a
typical pulse with a maximum field of $B=50$\,T occurring in $10$\,ms,
the field sweep rate can reach $\sim 5000$\,T/s. In this demanding
environment, fast optical techniques which are immune to
electromagnetic interference have significant advantages.

We first review the established methods of measuring magnetostriction in
pulsed fields, via capacitance dilatometry or resistive
foil strain gauges. 

Capacitance dilatometry is the standard technique used for both
thermal expansion and magnetostriction measurements in steady
fields. The main advantages are that the sample is under only a weak
uniaxial stress (applied via springs to keep the capacitor plates in
contact with the sample), and that the sensitivity increases the
closer the capacitor plates can be approached, while kept in a
parallel configuration. Using this technique, the minimum relative
change in sample length, $L$, that can be resolved is $\Delta L/L \sim
10^{-5}$ at 10--20\,kHz in pulsed fields\cite{doerr08} (or $\sim
10^{-7}/\sqrt{\text{Hz}}$). Such dilatometers are sensitive to
vibration, since the plates are only coupled to each other through the
sample, and to electrical noise in the measurement circuit. The
limited space inside pulsed field magnets acts as a further constraint
on the sensitivity, as the capacitor plates cannot be above a certain size.

Resistive foil gauges bonded directly to the sample have the benefit
of relative immunity from vibrations, but suffer from strong
magnetoresistance which must be calibrated out or well matched. Their
sensitivity is also quite low. For a 1\,k$\Omega$ strain
gauge a resistance change of 2\,m$\Omega$ must be detected to
resolve strains of $10^{-6}$. The best resolution so far reported
with this technique is $5 \times 10^{-6}$ in pulsed
fields.\cite{zaragoza06} One further disadvantage is the need to
bond strain gauges to the sample surface. The resulting 
inhomogeneous stress field can affect the response of softer
samples. On the other hand, this confers relative immunity from
vibration.

An electrically equivalent, but potentially more sensitive technique
is to use the the deflection of a piezoresistive cantilever to measure the
magnetostriction. This has been recently demonstrated in steady and
pulsed high fields.\cite{park09} Minimal stress is applied to the sample from contact
with the thin cantilever. The sensitivity is as yet unknown, however.

Common to all three techniques described above is the need for an
electrical bridge circuit to measure either resistance or
capacitance. The metallic components of the bridge circuit, such as
wiring and capacitor plates, are subject to self-heating and motion in
the rapidly varying magnetic field. In general, these cause both noise
and a background signal which can be difficult to subtract from the output.

In contrast, optical fibre strain gauges --- or Fibre Bragg Gratings
(FBGs) --- have the significant advantage of immunity to
electromagnetic interference. In addition they combine some of the key
advantages of the technologies described above. They can be bonded
directly to samples in the same way as resistive foil strain gauges,
or can be incorporated into a dilatometer design where they are used
both as spring and sensing element.

An FBG is based on a modulation of the refractive index in the core of
an optical fibre over some length. The period of this modulation
determines the wavelength of light reflected (the Bragg wavelength),
while the depth of modulation controls the reflectivity. Longer FBGs
have narrower linewidths.\cite{othonos97} FBGs have several
applications in the fields of telecommunications and sensors, and are
a quite mature (and hence cost effective and widely available)
technology.

In this article, we discuss the implementation of FBG strain gauge
dilatometry at cryogenic temperatures in pulsed high magnetic
fields and show that the resolution and interrogation rate are
limited only by mechanical noise and the measurement system
respectively. We discuss the ways in which strain can be transmitted
from the sample under investigation to the FBG. We describe a high
resolution, high speed measurement system capable of resolving changes
in strain of $\sim 10^{-9}/\sqrt{\text{Hz}}$ in pulsed fields.

\section*{Methods}

The Bragg wavelength, $\lambda_B$, of an FBG is related to the
refractive index of the optical fibre, $n$, and the pitch of the
grating, $\Lambda$:
\begin{equation}
\lambda_B = 2n\Lambda
\end{equation}

The strain and temperature dependence of $\lambda_B$ are given by:
\begin{equation}
\begin{split}
    \frac{\Delta\lambda_B}{\lambda_B} &=
    \biggl( 1 - \frac{n^2}{2}(P_{12} - \nu(P_{11}+P_{12}))\biggr)
    \frac{\Delta L}{L} \\
    &+ \biggl(\alpha(T) + \frac{1}{n}\frac{dn}{dT}\biggr)
    \Delta T
\end{split}
\end{equation}

The strain response depends on the collection of terms
$1-\frac{n^2}{2}(P_{12} - \nu(P_{11}+P_{12})) \approx
0.76$,\cite{kersey97} which has minimal temperature dependence from
2--300\,K for FBGs in silica fibre.\cite{james02} $P_{11}$ and
$P_{12}$ are components of the strain-optic tensor and $\nu$ is the
Poisson ratio. The strain response of an FBG is thus linear with the
same constant of proportionality at all temperatures, and no
calibration is required.

The temperature dependence is controlled by the linear thermal
expansion of the optical fibre, $\alpha(T)$, and the change in
refractive index with temperature. Both of these become very small at
$^4$He temperatures in silica fibres.\cite{reid98}

The change in $\lambda_B$ in high magnetic fields has been calculated
for the case of polarised light,\cite{diego04} and would lead to a
shift of around 0.1\,nm for an FBG with $\lambda_B=$1550\,nm in silica
at 100\,T. For the case of unpolarised light, as is used in typical
interrogation systems, $\lambda_B$ is not expected to change with magnetic
field.

\begin{figure}
  \includegraphics[width=0.47\textwidth]{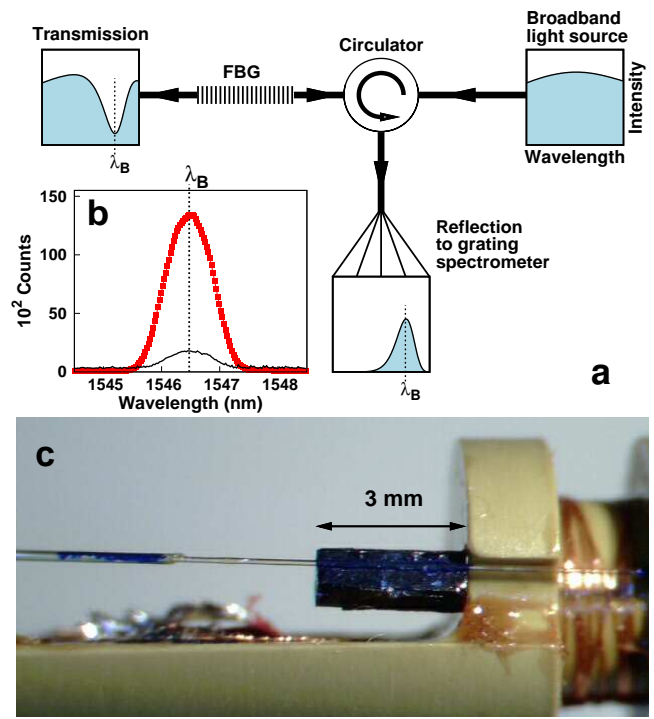}
  \caption{{\em a)} Optical circuit used for
    interrogation of the FBG. The boxes show schematic spectra of the
    light at each point in the circuit. {\em b)} The reflection
    spectrum for the FBG pictured above at 5.4\,K in
    zero field (red squares). The r.m.s. noise in each pixel is shown multiplied by
    100 (black line). {\em c)} Optical fibre containing 2\,mm FBG glued to a
    single crystal of cubic GdSb with cyanoacrylate. The sample
    dimensions are $3 \times 1.5 \times 1.5$\,mm$^3$. The sample is glued
    to a flat surface perpendicular to the optical fibre and sample
    $a$-axis using GE varnish.  This arrangement is appropriate for
    measuring $\Delta L || B || a$, where $B$ is the applied magnetic
    field. The measurement probe on which the sample is mounted is
    8\,mm in diameter, suitable for the smallest bore magnets. The
    optical fibre runs to the top of the probe where it passes through
    a vacuum feedthrough.}
  \label{fig:glued}
\end{figure}

We have chosen to work with FBGs written into the core of single-mode
telecommunications optic fibre with $\lambda_B = 1550$\,nm
at room temperature. Since this is an important telecoms band,
interrogation equipment and custom FBGs are readily available. The
expected shift of $\lambda_B$ is 1.2\,pm for 10$^{-6}$ strain. 

The fibre we use has a core diameter of 9\,$\mu$m and a cladding
diameter of 125\,$\mu$m. On top of this there is a protective plastic
coating.  The choice of fibre coating is important for strain-related
applications. The mechanical contact between sensor and sample must
accurately transmit the strain.  Polyimide is a good choice as it
sticks to the fibre strongly and there are many epoxies suitable for
fixing the FBGs to the test material. Including coating, the FBGs that
we use are $\sim$0.16\,mm in diameter and either 1 or 2\,mm long. This
is smaller than the majority of commercially available resistive foil
strain gauges.  A maximum strain of around $10^{-2}$ can be applied to
an FBG before the response becomes non-linear or the fibre
breaks.\cite{smartfibres} This is easily large enough for most strain
measurements.

There are then many choices of interrogation systems to use,
operating on diverse principles. For a good review, see
Ref.~\onlinecite{kersey97}. For measurements in pulsed fields a fast
technique ($>$\,10kHz) is necessary, whereas absolute accuracy and
long-term stability are less important.

We focus our attention on multi-channel spectrometry, which allows us
to detect the complete lineshape of the FBG reflection. If the
lineshape is distorted when we affix the FBG to the sample, we can
tell if the FBG is under homogeneous stress, {\em i.e.} if it is
uniformly coupled to the sample along its length.

Our high-resolution FBG interrogation system uses the fastest
currently available 1024-pixel InGaAs line array camera sensitive in
the wavelength range 900 to 1700\,nm, which can read full spectra at
47\,kHz.\cite{sensorsunlimited} The signal-to-noise ratio (SNR) is
specified at 5300:1 at full scale where readout and shot noise
dominate. Mounted on a fibre-coupled research spectrometer of focal
length 550\,mm with a 950\,l/mm diffraction grating, the 25\,$\mu$m
pixel pitch results in a dispersion, $\Delta\lambda$, of
0.027\,nm/pixel at 1550\,nm.

To illuminate the FBG we use a superluminescent-diode-based
broadband source with an output power of 60\,mW in the wavelength
range 1525-1565\,nm.\cite{denselight} The optical circuit is illustrated in
Figure~\ref{fig:glued}a. When the source characteristic is measured
directly using the spectrometer and camera described above, the SNR
reaches 800:1 for full-scale illumination at the shortest integration
time of 20.4\,$\mu$s. This is the dominant source of noise in the measurement.

A typical reflection spectrum for a 2\,mm FBG is shown in
Figure~\ref{fig:glued}b. The uncertainty in the measurement of
$\lambda_B$ depends on the way it is calculated from the returned
spectra. The standard technique is to calculate the centre of mass of
the reflected peak using:
\begin{equation}
  \lambda_B =\frac{\Sigma_i \lambda_i S_i}{\Sigma_i S_i}
\label{eqn:lambdab}
\end{equation}
where $\lambda_i$ is the wavelength at the centre of the $i$-th pixel
and $S_i$ is the corresponding signal amplitude. For a peak whose full
width covers $m$ pixels, the r.m.s. error in $\lambda_B$ is
approximately:
\begin{equation}
  \sigma_{\lambda_B} \approx \frac{m^{1/2}\Delta\lambda}{6\mathrm{ SNR}} \qquad (m \gg 1)
  \label{eqn:error}
\end{equation}
which has been obtained by a simulation excluding the effects of
camera readout noise. Our 2\,mm FBGs have a full width at half-maximum
(FWHM) of $1.0 \pm 0.1$\,nm and $m\sim 80$. For the parameters quoted
above, this should result in a strain resolution of $\Delta L/L \sim 5
\times 10^{-8}$ per reading, or $\sim 2\times
10^{-10}/\sqrt{\text{Hz}}$.

One pitfall of this approach is that the incident light power must be
increased as the rate of interrogation increases, since the detector
must be saturated in a shorter time to achieve the optimal
SNR. Fortunately, this power is not dissipated at the sample, rather
it is lost at the end of the fibre. If the cryostat is wider than the
minimum bend diameter of the fibre (10\,mm), its end can be positioned
back at the warm end of the cryostat where extra heat input is
irrelevant.  This means that FBGs can be interrogated at very low
temperatures without self-heating, making them suitable for operation
in $^3$He or dilution refrigerators. This is an improvement on
electrical measurements, where the excitation must be reduced at low
temperatures to avoid self heating, which necessarily reduces the SNR.

Coupling the FBG to the sample can be done in two ways. Our primary
consideration is that the experiment should fit inside the various
cryostats used for pulsed field measurements, the smallest of which
has an internal diameter of 8\,mm.

In the simplest case, for a regular sample with a flat surface parallel to
the direction of interest, the FBG can be bonded directly to the
sample using an appropriate epoxy.\cite{hbmepoxy} The considerations
are the same as for resisitive foil strain gauges: what curing
conditions (temperature and pressure) can be tolerated by the sample and FBG,
how good is the bond, how accurately can the FBG be aligned along a
crystalline axis? Of particular concern is that the sample can
experience an inhomogeneous stress due to the epoxy, which therefore
distorts the results.

To estimate if this is significant, consider that the Young's modulus
of silica is around 70\,GPa. This is similar to many metals. As long
as the sample has a much larger cross-sectional area than the optical
fibre, the presence of the fibre should not have a significant effect.

The advantages of direct bonding of the FBG to the sample surface are
compactness and less sensitivity to mechanical vibrations, which are
both concerns when working in small bore pulsed magnets. We have also
substituted cyanoacrylate (superglue) in place of a specialised epoxy,
with some success. This cures quickly at room temperature and is
soluble in acetone, allowing both sample and FBG to be
recovered. Figure~\ref{fig:glued}c shows a sample with an FBG glued
along its length mounted on a measurement probe.

The alternative is to couple the FBG and sample indirectly, through a
dilatometer. In such a dilatometer the sample is clamped into a
holder which is mechanically coupled to a free standing FBG sensor. In
this way the strain is directly transmitted to the sensor. This is
preferable for samples that are intolerant of applied stress, for
irregularly shaped samples, or for those which present a different
behaviour on the surface than in the bulk.

\section*{Results}

We have tested the FBG interrogation system described above in
pulsed magnetic fields at the Hochfeld-Magnetlabor Dresden.
We present results on single-crystal GdSb and GdSi. Owing to the lack of
crystal field effects ($L$~=~0 for Gd$^{3+}$), these two compounds are model
systems in which to study complex magnetic two-ion interactions. Both
reveal large field-induced magnetostriction.

\begin{figure}
\includegraphics[width=0.47\textwidth,clip]{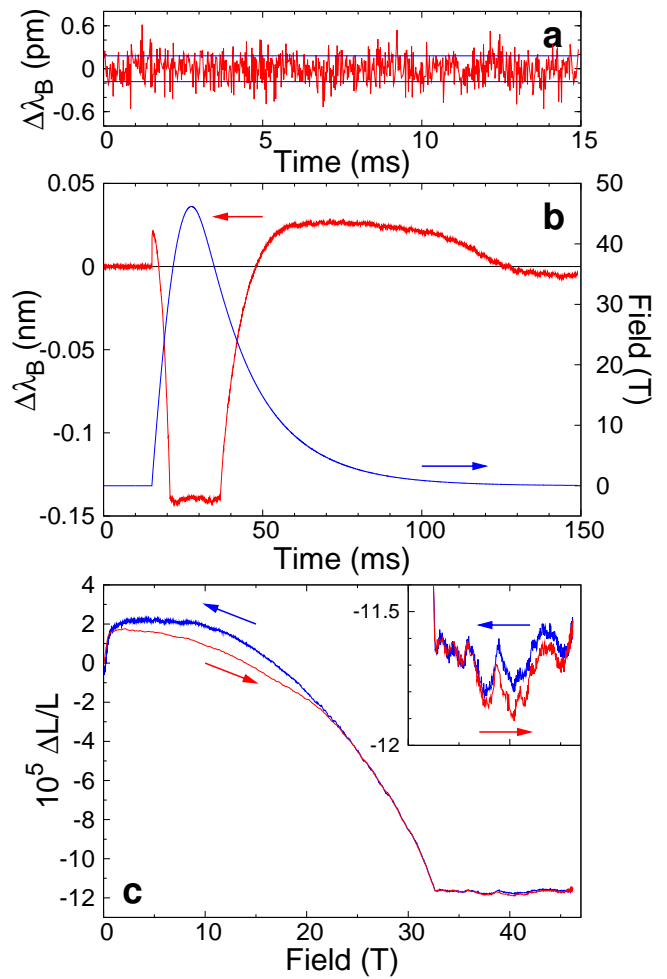}
  \caption{{\em a)} $\Delta\lambda_B$ over time in zero field as
    determined by Equation~\ref{eqn:lambdab}, for the FBG glued to the
    single crystal of GdSb at 5.4\,K, as shown in
    Figure~\ref{fig:glued}c (red line). The two horizontal lines are at
    $\pm \sigma_{\lambda_B} = 0.18$\,pm.  {\em b)} When exposed
    to the magnetic field (blue line), $\Delta\lambda_B$ recovers to
    nearly the same value after the field returns to zero.  {\em
      c)} The relative length change $\Delta L/L$ as a function
    of magnetic field determined from the data above. The field sweep
    direction is indicated by arrows. The increase below 2\,T is a
    spin-reorientation out of the normal antiferromagnetic state,
    while the sharp kink at $32.62 \pm 0.03$\,T corresponds to the
    point where the spins are fully ferromagnetically aligned. The up
    and down sweeps match extremely well at high fields. The signal
    is noisier as the field returns to zero where mechanical vibrations
    become stronger. The inset zooms in on the data above 30\,T, where
    small variations in $\Delta L/L$ at the 10$^{-6}$ scale are very
    well reproduced as field goes up and down, suggesting they are an
    intrinsic response of the sample. The random noise at high fields
    is around the same as in zero field.}
  \label{fig:gdsb}
\end{figure}

Figure~\ref{fig:gdsb} shows the response of the FBG on the GdSb single
crystal pictured in Fig.~\ref{fig:glued}c when exposed to the pulsed
magnetic field. GdSb is a member of the Gd monopnictide series Gd$X$
($X$=N, P, As, Sb and Bi) with a cubic NaCl-type crystal structure and
is characterized by a low carrier density. It orders
antiferromagnetically at 23\,K and has nearly no magnetic
anisotropy. The transition into the field-induced ferromagnetic state
proceeds over a wide field range up to 32.6\,T followed by a sharp kink
when saturation is reached. The magnetostriction measurement shows
this transition clearly. Despite the strong spin moment of
Gd$^{3+}$, the change in the magnetic moment is not accompanied by a
large lattice effect. The magnetostriction is in the order of
$10^{-4}$ only.

The r.m.s. variation of $\lambda_B$ over 15\,ms in zero field is shown
in Figure~\ref{fig:gdsb}a
to be $\sigma_{\lambda_B} = 0.18$\,pm, which corresponds to a strain
resolution of $1.5 \times 10^{-7}$. This is three times less than
the (optimistic) prediction of Equation~\ref{eqn:error}, and we
attribute the difference to a combination of camera readout noise and
mechanical vibration.

The noise at high fields is similar to that at zero field. There is
some evidence that the sample is heated by the rapidly changing
magnetic field, as the value of $\lambda_B$ immediately after the
pulse is slightly different to that before. This can be seen in
Figure~\ref{fig:gdsb}b. Since the thermal expansion intrinsic
to the optical fibre is negligble below 40\,K, it does not contribute
significantly to this shift.



\begin{figure}
  \includegraphics[width=0.47\textwidth,clip]{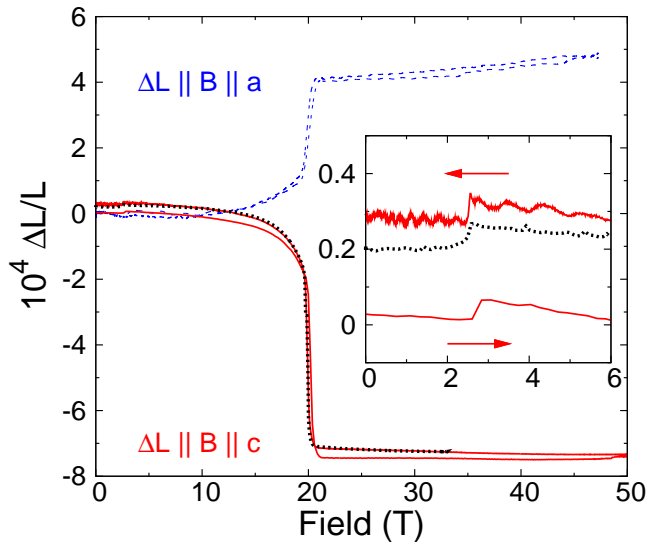}
  \caption{Relative length change $\Delta L/L$ of GdSi at 4.8\,K
    measured by 1\,mm FBG glued to the sample surface with
    cyanoacrylate (red solid line) as a function of magnetic field
    for $\Delta L || B || c$. The acquisition rate was 30\,kHz. The
    data is nearly identical to that taken in a 33\,T resisitive
    magnet on the same sample at 4\,K (black dotted line), which has been
    scaled by +5\% and offset to match the downsweep of the FBG
    data. The inset shows that the small transition of amplitude $\sim
    10^{-5}$ at 2.5\,T is clear in both measurements, indicating the
    quality of the pulsed field data. For comparison, data taken by
    capacitance dilatometry at 5\,K with an excitation frequencz of 10\,kHz (blue dashed
    line)\cite{doerr08} on the same sample but with $\Delta L ||
    B || a$, is noisier and has a large drift at high field where
    no further length change is expected as the magnetisation is saturated.}
  \label{fig:gdsi}
\end{figure}

The up- and downsweep data match extremely well above 20\,T when
plotted as a function of magnetic field Figure~\ref{fig:gdsb}c. Hence
any self-heating effect must be limited to long times or low fields.
More noise can be seen towards the end of the pulse (see also the
inset to Figure~\ref{fig:gdsi}). This may be due to mechanical
vibrations caused by the rapid boiloff of the liquid cryogens in the
system during the pulse.

In order to compare the performance FBG-based magnetostriction
measurements directly with established techniques, we have measured the same
single crystal of GdSi as was used in previous studies.
Figure~\ref{fig:gdsi} shows the magnetostriction of a single crystal
of GdSi measured by bonding a 1\,mm FBG to the sample with
cyanoacrylate. GdSi crystallizes in the orthorhombic FeB
structure. The magnetisation at low field shows anisotropic behaviour
due to the interplay between the localised $4f$-spins and the
conduction electrons.\cite{tung05} The magnetisation along the
$c$-axis merges with that measured along $b$ at about 3\,T. Note, that
this small anomaly is clearly visible in the magnetostriction curve
measured by FBG. At higher field the magnetisation becomes isotropic
and, after a spin-flop process, reaches saturation through a
first-order-like transition at 20\,T. Both processes are clearly
visible in the striction curve. The r.m.s. noise level in the GdSi
experiment is increased to about $\sigma_{\lambda_B} = 0.3$\,pm as the
shorter FBG has a wider reflection peak ($m\sim 160$). The shorter FBG
was required as the sample was less than 2\,mm long along the
$c$-axis.

The comparison of the FBG technique with high steady and pulsed field
capacitance dilatometry shown in Figure~\ref{fig:gdsi} is very
favourable. The high bandwidth allows sharp features to be easily
resolved, even on the fast upsweep, while the noise level approaches
the steady field results. The FBG signal at 4.8\,K is about 5\% smaller than the steady
field measurements at 4\,K. Assuming that the steady field measurements are authoritative, this
suggests that the transmission of strain from the sample to the FBG is
at worst 95\%, which is extremely good. The other possibility is that
the slight difference in sample temperature leads to this small
quantitative difference in the magnetostriction.

\section*{Conclusions}

We have demonstrated that FBGs are a very suitable technology for
high-accuracy magnetoelastic investigations in small samples
under the many simultaneous strong constraints imposed by low
temperatures and the highest avalable magnetic fields. 

We have developed an FBG interrogation system capable of determining
$\lambda_B$ with an r.m.s. error $\sigma_{\lambda_B}=0.18$\,pm and a
bandwidth of 47\,kHz. We have used this to measure the
magnetostriction of small samples glued straightforwardly to
FBGs, at cryogenic temperatures in the hostile environment generated
by pulsed magnetic fields. We are capable of resolving relative length
changes, $\Delta L/L$, as small as $3\times 10^{-7}$ (twice the
r.m.s. error in interrogation) on a timescale of 20\,$\mu$s, or
$1.4\times 10^{-9}/\sqrt{\text{Hz}}$. This is nearly two orders of magnitude
better than was previously possible by other techniques in
non-destructive high magnetic fields. This increase in resolution
greatly broadens the range of materials whose magnetostriction
can be investigated in pulsed magnetic fields.

The FBG based technique has the significant advantage of immunity from
electrical noise, and is furthermore quite simple to implement. Since
there are no metallic parts, there are no self-heating effects in the
magnetic field. No calibration is required. We have shown that the strain is well
transmitted from the sample to the FBG.

It should therefore be easy to extend the technique to measure strain
transverse to the applied field as well. This can be done
simultaneously with a longitudinal measurement using a single fibre
containing two FBGs with different $\lambda_B$ glued to the sample in
different orientations. No further equipment is required. Similarly,
multiple samples can be measured simultaneously using a single fibre,
the only limitations being space in the cryostat at the centre of the
field and the spectral range of the spectrometer.

The technique is also compatible with very low temperatures
($<$1\,K), as in principle no heat is dissipated at the cold end
of the cryostat. Inserting the optical fibre into a high-pressure
piston-cylinder type clamp cell to measure magnetostriction under
pressure in pulsed fields is also a promising possibility. Much
potential exists to extend the utility of FBGs further.

\begin{acknowledgments}
This work was supported by the MPG Research Initiative ``Materials
Science and Condensed Matter Research at the Hochfeld-Magnetlabor
Dresden''. We would like to thank M.~Bartkowiak and S.W.~James for useful discussions and
F.~Tr\"{u}ller for technical support.
\end{acknowledgments}


\begin{thebibliography}{99}

\bibitem{doerr08} M.~Doerr, W.~Lorenz, T.~Neupert, M.~Loewenhaupt,
  N.V.~Kozlova, J.~Freudenberger, M.~Bartkowiak, E.~Kampert and
  M.~Rotter, {\em Rev. Sci. Instrum.} {\bf 79}, 063902 (2008).

\bibitem{zaragoza06} P.A. Algarabel, A. del Moral, C. Mart,
  D. Serrate and W. Tokarz {\em J. Phys.: Conference Series} {\bf 51}
  607 (2006).

\bibitem{park09} J.-H. Park, D. Graf, T. P. Murphy, G. M. Schmiedeshoff
and S. W. Tozer, {\em Rev. Sci. Instrum.} {\bf 80} 116101 (2009).

\bibitem{othonos97} Andreas Othonos {\em Rev. Sci. Instrum.} {\bf 68}
  4308 (1997).

\bibitem{kersey97} Alan D. Kersey, Michael A. Davis, Heather
  J. Patrick, Michel LeBlanc, K. P. Koo, C. G. Askins, M. A. Putnam,
  and E. Joseph Friebele {\em J. Lightwave Tech.} {\bf 15} 1442 (1997).

\bibitem{james02} Stephen W. James, Ralph P. Tatam, Andrew Twin, Mungo
  Morgan and Paul Noonan {\em Meas. Sci. Technol.} {\bf 13} 1535 (2002).

\bibitem{reid98} B. Reid and M. Ozcan {\em Opt. Eng.} {\bf 37} 237 (1998).

\bibitem{diego04} J.L. Arce-Diego, D. Pereda-Cubian and M.A. Muriel
  {\em J. Opt. A: Pure Appl. Opt.} {\bf 6} S45 (2004).

\bibitem{smartfibres} See for example: http://www.smartfibres.com/docs/SmartFBG.pdf

\bibitem{sensorsunlimited} Goodrich ISR (Sensors Unlimited) model
  SU1024LDH-1.7RT-0500/LC

\bibitem{denselight} Denselight Semiconductors model DL-ASE-CW-CSC183A.

\bibitem{hbmepoxy} For example, EP310-S from HBM GmbH.

\bibitem{tung05} L.~D.~Tung, M.~R.~Lees, G.~Balakrishnan,
D.~McK.~Paul, P.~Schobinger-Papamentellos, O.~Tegus, P.~E.~Brommer,
and K.~H.~Buschow {\em Phys. Rev. B} {\bf 71} 144410 (2005).

\end{thebibliography}
\end{document}